\documentclass{aastex}
\usepackage{emulateapj5}

\usepackage{subfigure}

\slugcomment{Submitted to Astronomical J.}

\shorttitle{The PPMXL catalog}
\shortauthors{Roeser et al.}

\makeatletter

\newenvironment{figurehere}
  {\def\@captype{figure}}
  {}
\makeatother

\begin{document}


\title{The PPMXL catalog of positions and proper motions on the ICRS.\\
    Combining USNO-B1.0 and 2MASS.}
    

\author{S. Roeser, M. Demleitner and E. Schilbach}
\affil{Astronomisches Rechen-Institut, Zentrum f\"ur Astronomie der
    Universit\"at Heidelberg, M\"onchhofstr. 12-14, 69120 Heidelberg,
    Germany}



\begin{abstract}
USNO-B1.0 and 2MASS are the most widely used full-sky surveys.
However, 2MASS has no proper motions at all, and USNO-B1.0
published only relative, not absolute (i.e. on ICRS) proper motions.
We performed a new determination of mean positions and proper motions
on the ICRS system by combining USNO-B1.0 and 2MASS astrometry.
This catalog is called PPMXL\footnote{VO-access to the catalog
is possible via http://vo.uni-hd.de/ppmxl}
, and it aims to be complete from the
brightest stars down to about $V \approx 20$ full-sky.
PPMXL contains about 900 million objects,
some 410 million with 2MASS photometry, and is the largest
collection of ICRS proper motions at present.
As representative for the ICRS we chose PPMX. The recently released
UCAC3 could not be used because we found
plate-dependent distortions in its proper motion system
north of -20$^\circ$ declination. UCAC3 served as an intermediate system for
$\delta \leq -20^\circ$.
The resulting typical individual mean errors of the proper motions
range from 4 mas/y to more than 10 mas/y depending on observational
history. The mean errors of positions at epoch 2000.0 are 80 to 120 mas,
if 2MASS astrometry could be used, 150 to 300 mas else.
We also give correction tables to convert USNO-B1.0 observations
of e.g. minor planets to the ICRS system. 
\end{abstract}





\section{Introduction}
According to IAU Resolution B2 of the XXIIIrd General Assembly (1997), the Hipparcos
catalog \citep{1997yCat.1239....0E} is the primary realisation of the International
Celestial Reference System (ICRS)
at optical wavelengths. Its first and basic extension to higher star densities and fainter limiting
magnitudes is Tycho-2 \citep{2000A&A...355L..27H},
based on observations of the Tycho experiment onboard the ESA-Hipparcos satellite.
The early epoch observations of Tycho-2 were taken from new reductions \citep{1998AJ....115.1212U}
of the
observations made for the Astrographic Catalog  and 143 other ground-based astrometric
catalogs.
Tycho-2 contains about 2.5 million stars and is 90 percent complete down to
$V$ = 11.5. 
\citet{2008A&A...488..401R}  published the PPMX catalog of positions and proper motions
of 18 million stars with limiting magnitude around 15 in a red band. The typical 
accuracy of the proper motions is about 2 mas/y for 4.5 million stars with first epoch
in the Astrographic catalog and about 10 mas/y for all other stars. 

Very recently, the UCAC3 catalog \citep{2009yCat.1315....0Z}
was released. UCAC3 is based on a new full-sky astrometric survey
made in the years 1998 to 2004. The catalog contains some 100 million stars down to $r_U$ = 16 mag.

The largest catalog in the optical regime is USNO-B1.0 \citep{2003AJ....125..984M}
with more than one billion objects. However, USNO-B1.0
is not in the system of ICRS; it contains relative, not absolute proper
motions  \citep[see][]{2003AJ....125..984M}. A comparison of USNO-B1.0 and
PPMX performed in the present work yielded systematic differences (in areas of
square degrees) of up to 15 mas/y in proper motion and up to
0.6 arcseconds in positions at epoch 2000.0.

The Two Micron All Sky Survey \citep{2006AJ....131.1163S}, 2MASS, is a complete Sky-Survey in
the J, H and K$_s$ bands performed in the years from 1997 to 2001.
The Point Source Catalog of about 471 million entries is
also a source of accurate
astrometric positions, but contains no proper motions.

For kinematical studies in the Milky Way a catalog of proper motions in the ICRS system
and with a well-defined completeness limit is indispensable. PPMX fulfills
this requirement, but with only 18 million stars it is rather small.
USNO-B1.0 with inertial proper motions would be a big step forward.
When SDSS observations became available, inertial proper motions have been constructed
from a combination of SDSS with USNO-B1.0 using SDSS galaxies as reference
\citep{2004AJ....127.3034M,2008AJ....136..895M,2004ApJS..152..103G}.
This is, of course, restricted to the SDSS
part of the sky.

The fact that USNO-B1.0 is not on ICRS creates a problem for minor planet 
observers. Right ascensions and declinations based on USNO-B1.0, when combined
with older epoch observations on ICRS may cause biases in orbit determinations
of minor planets. This is of great importance for fly-by manoeuvers of interplanetary
spacecraft \citep{2009DDA....40.1704C}.

Combining USNO-B1.0 with 2MASS is such an obvious idea that it is already mentioned
by \citet{2003AJ....125..984M}.
In the following  we first give a schematic overview of the entire
procedure(section \ref{over}), then we describe in detail the initial steps to {\em coarsely} put
USNO-B1.0 to the ICRS (section \ref{prel}). This is followed by a description the combination with 2MASS observations
(section \ref{lsqadj}), the details of the 
construction of the system of positions and proper motions on ICRS (section \ref{system}), and we close
with an overview of the properties of the catalog (section \ref{finalcat}).
Our approach can be considered an affordable effort to put USNO-B1.0 onto ICRS, and
improve the individual proper motions by inclusion of 2MASS. A sophisticated re-reduction
of all the material might deliver superior results provided that a better reference catalog
were available before Gaia.

\section{Overview on the reduction procedure}~\label{over}

In this section we give an executice summary of all the steps that lead to the construction
of the catalogue:

Step 1: Reconstruction of the individual observations that went into USNO-B1.0. Result:
$\alpha,\delta$ and epoch.

Step 2: Identification of USNO-B1.0 stars in a given ''field'' (see next section) with PPMX and 
determination of corrections $\Delta\alpha$, $\Delta\delta$ from the mean offset per field. Result:
$\alpha,\delta$ on the PPMX system on scales equal or larger than the field size.

Step 3: Cross-matching the USNO-B1.0 stars with 2MASS using a cone-search with radius 2 arcsec.
Result: $\alpha,\delta$ and epoch from 2MASS for the given USNO-B1.0 star.

Step 4: Attributing weights to each observation and  weighted least-squares adjustment for
each star. Result: new positions and proper motions of each star. 

Step 5: The positions and proper motions of all stars 
with $K_s$-magnitudes between 12 and 13 from step 4 represent the preliminary system PS1.
Cross-matching the stars in PS1 with UCAC3. South of -20$^\circ$ declination,
adding systematic differences UCAC3-PS1 in proper motions on a grid with
0.25$\times$0.25 deg$^2$ bins with a 3$\times$3 bin moving average filter.
The proper motions of PS1 north of -20$^\circ$ declination are left unchanged. This system
is called PS2.

Step 6: Cross-matching the stars in PS2 with PPMX
and adding systematic differences PPMX-PS2 in proper motions on a grid with
1$\times$1 deg$^2$ bins with a 3$\times$3 bin moving average filter. This is
called PS3 and completes the proper motion system of PPMXL.

Step 7: Putting stars without 2MASS observations onto the system PS3.
Make a least-squares adjustment (as in step 4) giving all 2MASS observations
weight zero. Compare the proper motions of all stars in PS2 with the proper
motions obtained in this step and add the systematic differences to a non-2MASS
star on a grid with
0.25$\times$0.25 deg$^2$ bins with a 3$\times$3 bin moving average filter.

Step 8: Putting the system of position 2000.0 onto the ICRS. 
Cross-matching the stars in PS2 with PPMX and determine the differences
in position at epoch 2000.0. Add systematic differences PPMX-PS2 in positions on a grid with
1$\times$1 deg$^2$ bins with a 3$\times$3 bin moving average filter.
Proceed analogously to Step 7 for the positions of non-2MASS stars.

\section{Initial steps}~\label{prel}

USNO-B1.0 is an impressive piece of work.  The individual positions and proper motions
are derived from up to 5 original observations. 
For each star, USNO-B1.0
publishes not only positions for the epoch 2000.0 and proper motions, but 
also gives additional information that enable to re-construct  all the original observations,
i.e. offsets in the x-coordinate (negative right ascension) and the y-coordinate (declination)
together with field and survey identifiers which allow to recover the observational epoch.
For the different surveys used for the construction of USNO-B1.0 see \citet{2003AJ....125..984M}.
Altogether USNO-B1.0 is divided into 7435 fields, and Table 3 of \citet{2003AJ....125..984M}
gives the necessary information to reconstruct all the individual observations that went into USNO-B1.0.
The epochs for the fields can mostly be found in http://www.nofs.navy.mil/data/fchpix/.
Additional epoch information was provided by Dave Monet.
Epochs and corrections
for all fields contributing to PPMXL can now be found at
http://vo.uni-hd.de/usnob/res/usnob/pc/form.

As astrometric reference USNO-B1.0 uses SPM and NPM (the Lick
Southern and Northern Proper Motions program),
which has the advantage of giving
a dense grid of reference stars on each field, but the disadvantage that the mean epoch
of SPM and NPM is about 1975. Also, the mean motion between the Schmidt plates of USNO-B1.0
and the SPM and NPM was set to zero in the least-squares adjustment.

Given the situation above, the re-constructed individual right ascensions and declinations
could not be combined immediately with 2MASS, because they are on different reference systems.
To overcome this,  we made the following assumption: as a dense reference catalog had been
used in each field (roughly corresponding to a Schmidt plate),
the offset of a field (at the field epoch) from ICRS can be described
to zeroth order by $\Delta\alpha$, $\Delta\delta$, the mean deviations in right ascension
and declination (at epoch) from a suitable reference
catalog on ICRS.
Since USNO-B1.0 gives all stars from Tycho-2 with their Tycho-2 entries,
these cannot be used directly as a reference. Instead, we used the stars in PPMX fainter
than the Tycho-2 limits  and cross-identified them with USNO-B1.0. Although the fainter
part of PPMX is based partly on the same surveys 
this approach is justified because PPMX as a whole is already on ICRS whereas USNO-B1.0
is not.

It turned out that this simple but straighforward approach yielded remarkably good
results north of -20$^\circ$ declination, proving that the plate
reductions using SPM and NPM were highly successful in these areas of the sky.
This completes steps 1 and 2 from the previous section.

In step 3
USNO-B1.0 was cross-matched with 2MASS using a cone-search with radius 2 arcsec.
The crossmatches were performed with in a PostgreSQLA database using
the q3c indexing scheme \citep{2006ASPC..351..735K}.
Double identifications are allowed; consequences thereof are discussed in section \ref{properties}.
After these steps USNO-B1.0 observations could be combined with the 2MASS observations.
The construction of
the final system, however, has been deferred to after the least-squares adjustment
of mean positions and proper motions.

\section{Least-squares adjustment of mean positions and proper motions}~\label{lsqadj}

Before entering a least-squares adjustment, individual weights $w_i = \sigma_{u.w.}^2/{\sigma_i}^2$ must 
be attributed to the observations of a star at epoch $T_i$ (Step 4).
The a-priori error of unit weight $\sigma_{u.w.}$ is arbitrarily set to 1 mas. The assignment of $\sigma_i$ is always
discussible. We attributed $\sigma_i = 230$~mas for each individual observation from USNO-B1.0.
\citet{2004AJ....127.3034M} give $\sigma_i = 120$~mas for an USNO-B1.0 position.
\citet{2003AJ....125..984M} however,
note that they found systematic offsets of up to 250~mas USNO-B1.0 positions when compared with 
the SDSS EDR.
Our weight corresponding to $\sigma_i = 230$~mas is therefore a reasonable estimate.
This weight has been chosen independent of the magnitude of the star.
If we assume that the accuracy of individual
observations in each survey in USNO-B1.0 deteriorates proportionally, then the resulting
proper motion itself is not affected, however the covariance matrix is. A deeper
investigation of magnitude dependent effects on individual accuracies in USNO-B1.0 was
beyond this work. 

The astrometric accuracy of 2MASS is 
80 mas (1 $\sigma$)  relative to the Hipparcos reference frame for $K_s < $ 14, and 
increases to 250 mas at $K_s = $ 16 \citep{2006AJ....131.1163S}. We used these values
for the $\sigma_i$ of a 2MASS observation.

The number of observations per star for the lsq-adjustment varies from 2 (one early and
one late epoch) to 6 (up to 5 epochs from the USNO-B1.0 and one from 2MASS).  
Given the notations above, we determine the resulting covariance matrix of the unknowns as

\begin{eqnarray}
\begin{tabular}{ c  }
$w  = \sum_{i=1,n}^{} w_i$ \hspace*{0.5cm}   $  w \overline{T} = \sum_{i=1,n}^{} w_i T_i $\\
\hspace*{0.5cm} \\
$w_{p.m.}  =
\sum_{i=1,n}^{} w_i (T_i-\overline{T})^2, $\\
 \end{tabular}
 \end{eqnarray}
where  $\overline{T}$ is the mean epoch, $w$ is the weight of the mean position, and 
$w_{p.m.}$ is the weight of the resulting proper motion. Because of the few degrees of freedom
($\leq$ 6 observations for 2 unknowns) in each lsq-adjustment
we did not determine an a posteriori error of unit weight for individual stars.
Hence, $ \sigma_{p} = w^{-1/2} $ and $ \sigma_{p.m.} = w_{p.m.}^{-1/2} $ are the
mean errors of
position and proper motion per coordinate, respectively.

The mean positions $\overline{x}$
and proper motions $\mu$ for each object are computed as

\begin{equation}
\overline{x} = \frac{\sum_{i=1,n}^{} w_i x_i}{\sum_{i=1,n}^{} w_i} \hspace*{0.6cm} \mu  = \frac{\sum_{i=1,n}^{} w_i x_i
(T_i-\overline{T})}{\sum_{i=1,n}^{} w_i  (T_i-\overline{T})^2}.
\end{equation}

Automatic tests for unduly large scatter among the measurements (based on the $\chi^2$ sum)
and automatic elimination of outliers 
(based on appropriately normalised individual residuals) were implemented.
All stars having $\chi^2$ sums beyond a certain significance limit, but still not
showing obvious outliers, were marked as ``problem cases'' and got a ``P'' flag
in the catalog.

\section{The system of positions and proper motions}~\label{system}

With the coarse systematic corrections described in section  \ref{prel} and the 
least-squares solution, a {\em preliminary catalog} of positions and proper
motions was constructed with only the final systematic corrections missing.

\subsection{The proper motion system}~\label{pmsystem}

To link a proper motion system of a sky survey (catalog) to the ICRS, two distinct approaches
are possible. Either the ''proper motions'' of extragalactic objects are forced to vanish,
this is the method adopted by \citet{2004AJ....127.3034M} or \citet{2004ApJS..152..103G} in their
reduction of SDSS proper motions, or, the optical representation of the ICRS, the Hipparcos catalog,
is extended to fainter magnitudes.
We chose the second alternative, as we did not try to identify point-source-like extragalactic
objects in USNO-B1.0. Also, at low galactic latitudes this method can hardly work.

\figurehere
\includegraphics[bb= 36 230 550 580,angle=0,width=0.45\textwidth,clip]{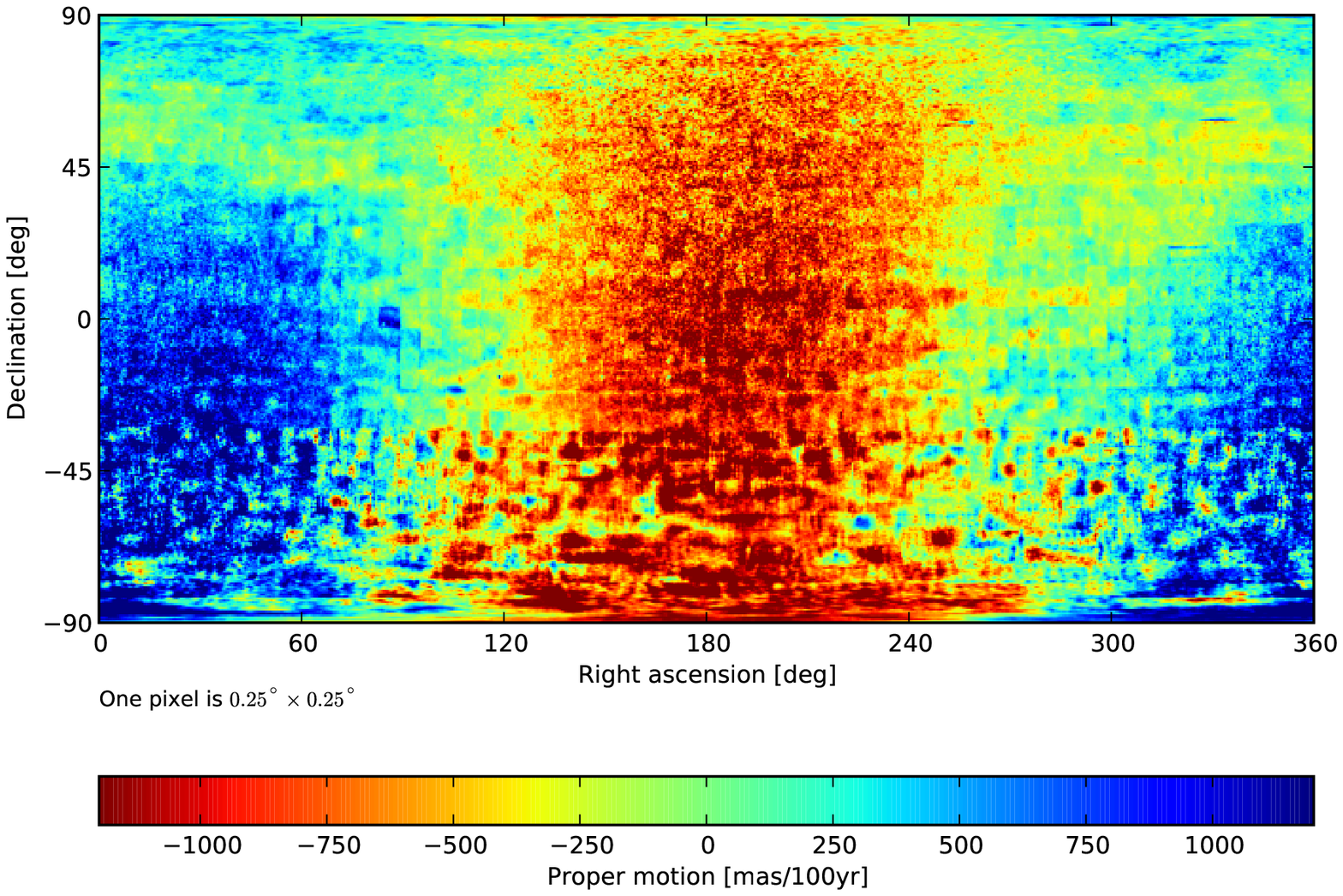}
\caption{Proper motions in right ascension $\mu_{\alpha}\cos\delta$
of the {\em preliminary system} PS1 plotted over right ascension and declination.
The plate structure of the underlying survey is clearly visible south of $\delta$ = -20$^\circ$.
Unless otherwise stated all the plots of this kind as a function of right ascension, declination refer to a grid with
0.25$^\circ\times$0.25$^\circ$ bins, and the quantities plotted are the averages from a 3$\times$3 bin
moving average filter.
}
\label{muealPS}
\vspace*{0.1cm}
The Hipparcos catalog itself is, of course, unsuited for the link because of its low spatial density
and its bright stars. The next obvious choice, Tycho-2, is  is unsuitable as well for reasons laid
out in section \ref{prel}.
This leaves the fainter part of the PPMX for the construction of the
link between the USNO-B1.0 proper motions and the ICRS.

\figurehere
\includegraphics[bb= 36 230 550 580,angle=0,width=0.45\textwidth,clip]{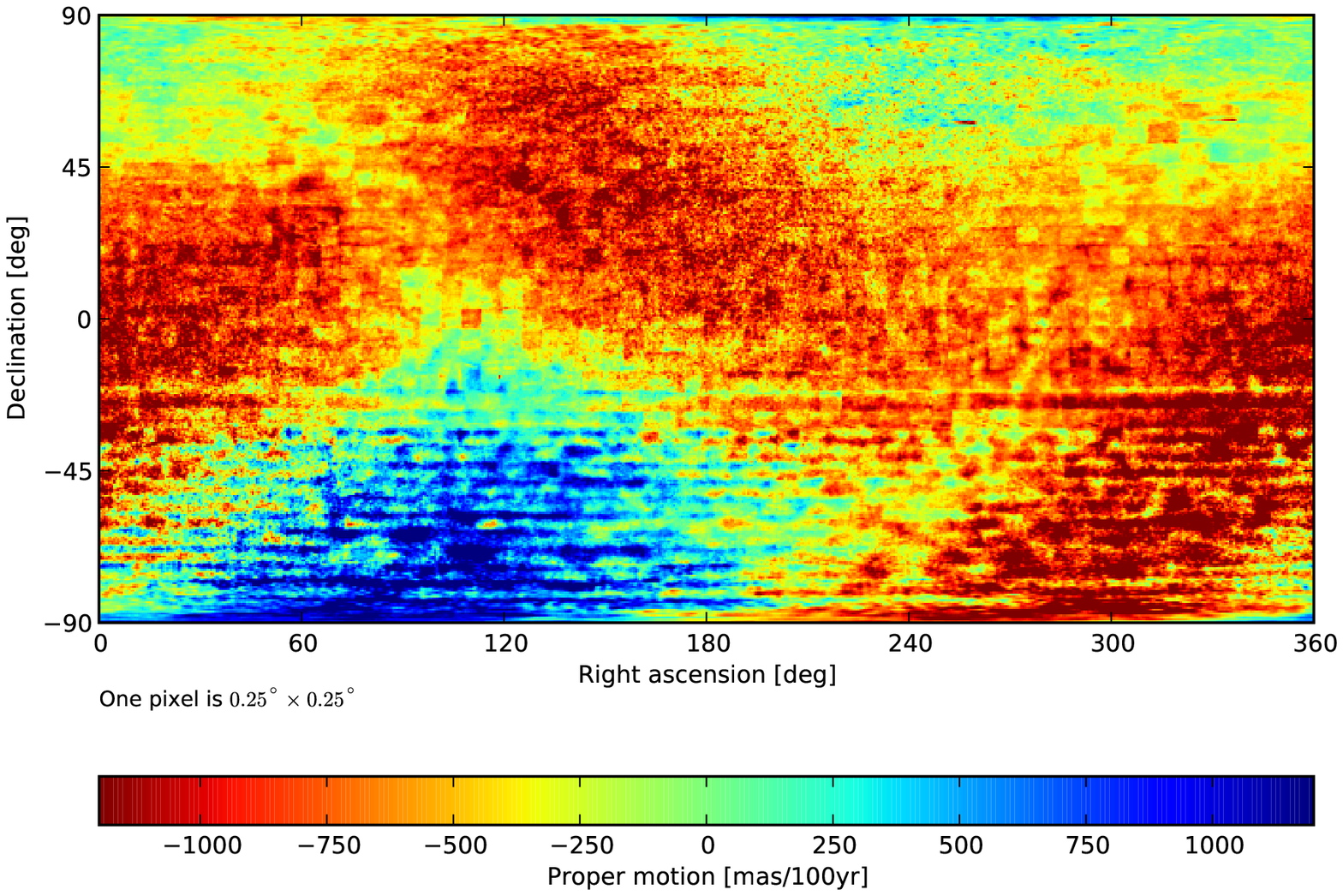}
\caption{Same as Fig.\ref{muealPS} but for the proper motions in declination $\mu_{\delta}$.
Again the system is distorted south of $\delta$ = -20$^\circ$.}
\label{muedelPS}
\vspace*{0.1cm}

Before we come to this comparison, let us note that the proper motion
system can be checked, to a certain degree, independent of a comparison with a reference catalog.
The proper motions
in an inertial reference system must, except for their peculiar motions,
on average only reflect the physical motions of the stars
in our Galaxy, i.e. the reflex of solar motion and the rotation of the Galaxy. Systematics
parallel to the axes of right ascension or declination must not appear, nor should  features be seen
representing the plate lay-out of a photographic survey.
In step 5 the stars in the {\em preliminary catalog} with
2MASS $K_s$-magnitudes between 12 and 13 have been chosen to represent the preliminary system (PS1).
Doing so, we get a fair overlap with the faint stars in PPMX.
The magnitudes 
in the visual range from USNO-B1.0 are not suited, because the magnitude system is very
inhomogeneous from plate to plate.

Figs. \ref{muealPS} and \ref{muedelPS} show the proper motions of the PS1
in the right ascension, declination plane in bins of 0.25$^\circ\times$0.25$^\circ$ averaged via a 3$\times$3 bin
moving average filter.
Surprisingly, the proper motion system is
remarkably smooth north of $-20^\circ$ declination. Only minor plate-dependent effects
can be seen. This is a hint that the authors of USNO-B1.0 achieved very satisfactory results 
in their plate reductions in this portion of the sky.
However, south of -20$^\circ$  declination, plate-dependent distortions of the
proper motion system are obvious and need to be corrected for.

\figurehere
\includegraphics[bb= 36 230 550 580,angle=0,width=0.45\textwidth,clip]{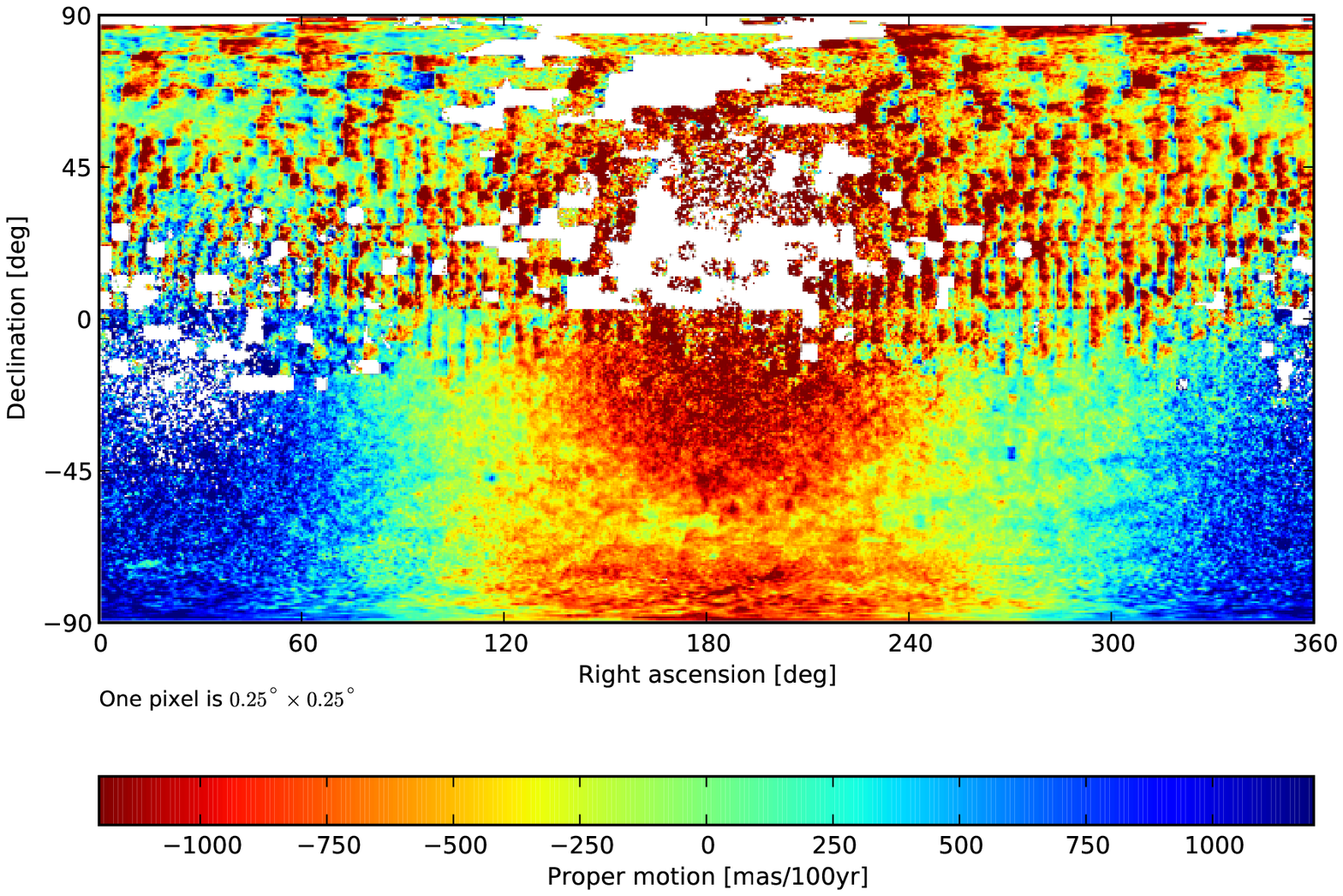}
\caption{Proper motions in right ascension $\mu_{\alpha}\cos\delta$
of UCAC3 in the magnitude range $r_U$ 14 to 15 plotted over right ascension and declination.
Empty areas stand for regions where no proper motions in this
magnitude range are available in UCAC3.
The proper motions of UCAC3 north of $\delta$ = -20$^\circ$
show strong plate dependent systematic distortions and cannot be used for Galactic kinematics.
However, the system south of $\delta$ = -20$^\circ$ is very well defined.
}\label{muealU3S}
\vspace*{0.1cm}

In August 2009, UCAC3 was released. UCAC3 contains some 100 million
stars and therefore goes deeper than PPMX, and, in principle, could be used for
the correction of PS1.
To characterise the UCAC3 system (U3S) of proper motions
we chose stars with $14<m_{r_U}<15$ from UCAC3.
This is well away both from the bright Tycho-2 stars and from the magnitude limit
of UCAC3.
  
Figs. \ref{muealU3S} and \ref{muedelU3S} show the proper motions of the U3S in bins of 0.25$^\circ\times$0.25$^\circ$ averaged via a 3$\times$3 bin
moving average filter.
White pixels show areas where UCAC3 (14 $ < r_U < $ 15) contains only stars
without proper motions. These areas
presumably largely coincide with those without first epochs in the
UCAC3 project.
We interpreted proper motion zero and negative errors on it (found for
about 9 million stars) as signifying null values in the corresponding
columns of UCAC3.

While the proper motion system is well-determined south of -20$^\circ$ declination,
it clearly shows unphysical effects in both coordinates north of -20$^\circ$, where
pattern-dependent proper motions occur with amplitudes
exceeding $\pm$ 12 mas/y. According to Zacharias (2009, private communication) the source
of the systematic effects comes from the Schmidt plates used for the first-epoch
positions, whereas in the south ($\delta < -20^\circ$) the plates from the SPM  could be used.   
These systematic errors in UCAC3 north of $\delta = -20^\circ$ are so large that UCAC3 cannot be considered
to be on the ICRS system, and the reader is advised to take care when using it for kinematic investigations
of our galaxy.

\figurehere
\includegraphics[bb= 36 230 550 580,angle=0,width=0.45\textwidth,clip]{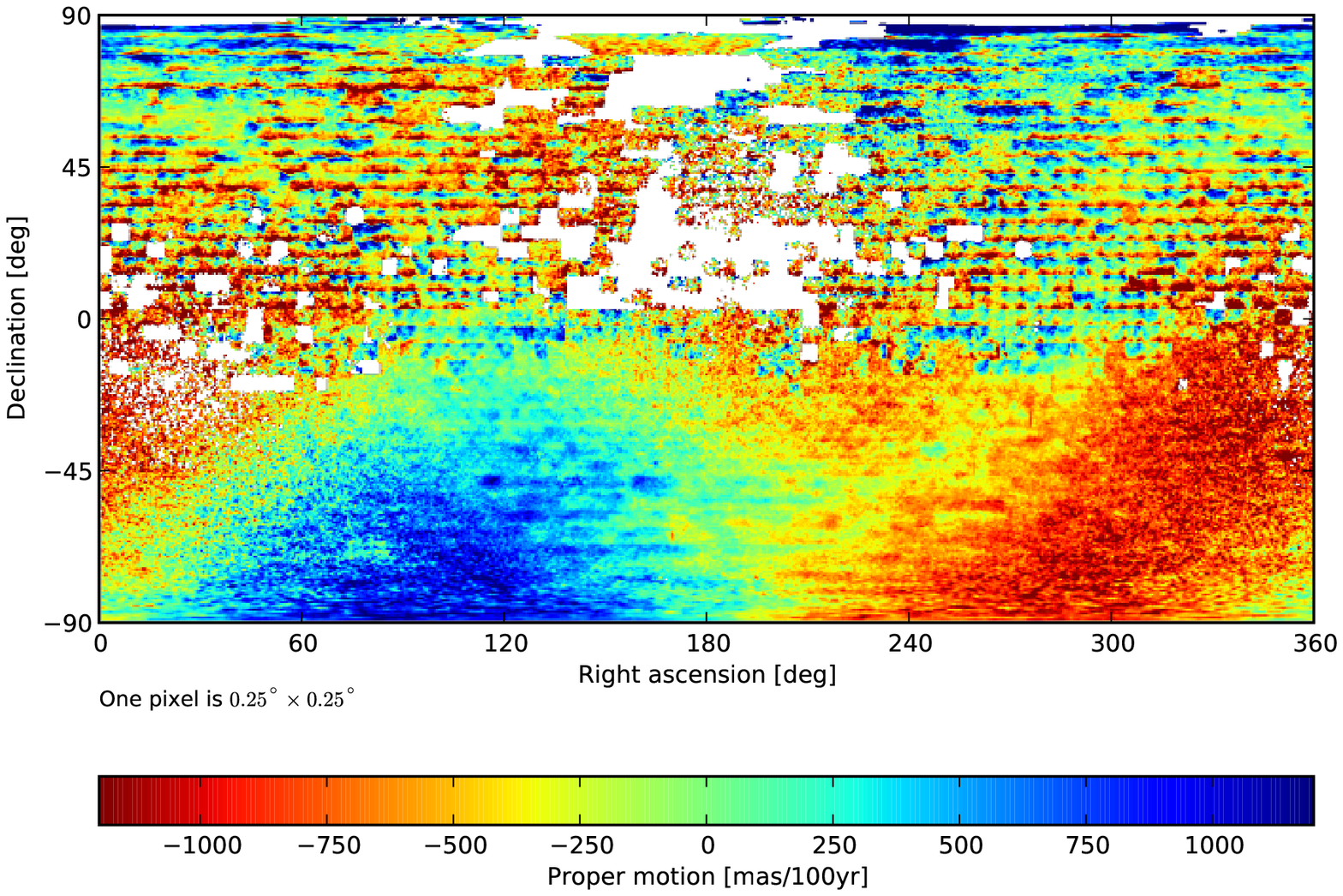}
\caption{Same as Fig.\ref{muealU3S} but for the proper motions in declination $\mu_{\delta}$.
Again UCAC3 proper motions should not be used north of $\delta$ = -20$^\circ$.
\label{muedelU3S}}
\vspace*{0.1cm}



To summarize the situation, it happens that we have
proper motion systems north and south of $\delta = -20^\circ$  which do not show conspicious
plate-dependent structure.
This
enabled us to construct an intermediate system as a combination of both, i.e the proper motions
of the PS1 are left unchanged north of $\delta = -20^\circ$, and differences U3S - PS1 are applied south
of $\delta = -20^\circ$  on
a grid with 0.25$\times$0.25 deg$^2$ bins with a 3$\times$3 bin moving average filter.
The resulting system is called PS2. This completes step 5.

Absence of plate-dependent distortions alone does not place a proper motion system onto the ICRS, the
link to the Hipparcos system is mandatory. In step 6, we chose 
PPMX as respresentative for Hipparcos, and we added systematic differences PPMX-PS2 to PS2 on
a grid with 1$\times$1 deg$^2$ bins with a 3$\times$3 bin moving average filter. This lay-out was chosen
to have enough PPMX stars for the link. Their number varies between 1000 and 5000, with a few areas
containing only 500 stars per bin. 
This link completes
the proper motion system of PPMXL. The resulting system PS3 is shown in Figs. \ref{muealfinal} and \ref{muedelfinal}.
The link to PPMX introduced a negative systematic in Fig. \ref{muealfinal} compared to Fig. \ref{muealPS}
at +40$^\circ$ declination. This depression is already inherent in Tycho-2 as shown by \citet{2008A&A...488..401R}.

So far, we only discussed the proper motion system of PPMXL for stars having 2MASS observations.
To ensure that objects in PPMXL beyond the 2MASS
limits are on the same proper motion system, we made a least-squares adjustment giving
weight zero to 2MASS observations (Step 7).
Then, we compared this proper motion system with PS3
and applied the resulting
corrections on
a grid with 0.25$\times$0.25 deg$^2$ bins with a 3$\times$3 bin moving average filter.
In other words, for the stars in PS3 we derived positions and proper motions with and without
including 2MASS observations and, so, corrected the stars that happen to have no 2MASS observations.

It is well known that magnitude- and colour-dependent systematic errors occur in astrometric
observations, be they photographic, taken with CCDs or even with photoelectric meridian circles.
In the case of USNO-B1.0 or PPMXL they are hard to be detected, as there is no independent
reference on ICRS at fainter magnitudes, e.g.  
at ``blue'' or ``red'' magnitudes between 19 and 20. 
 Also, these stars
have kinematics different from the bright stars in solar reflex and galactic motion.
So, to a certain extent is it difficult at present to
distinguish between systematic errors and different kinematics.
There is only one thing one can check:
the plate structure of the underlying surveys { \em must not} be seen. As an example we show in
Fig. \ref{muedelfaint} the proper motion system of PPMXL in the I-band of USNO-B1.0 in
magnitudes 18 $<$ I $<$ 19. We chose the I-band here as respresentative, because 
the I-magnitudes are more homogenous over the full sky than are the ''blue'' or ''red'' magnitudes.

\figurehere
\includegraphics[bb= 36 230 550 580,angle=0,width=0.45\textwidth,clip]{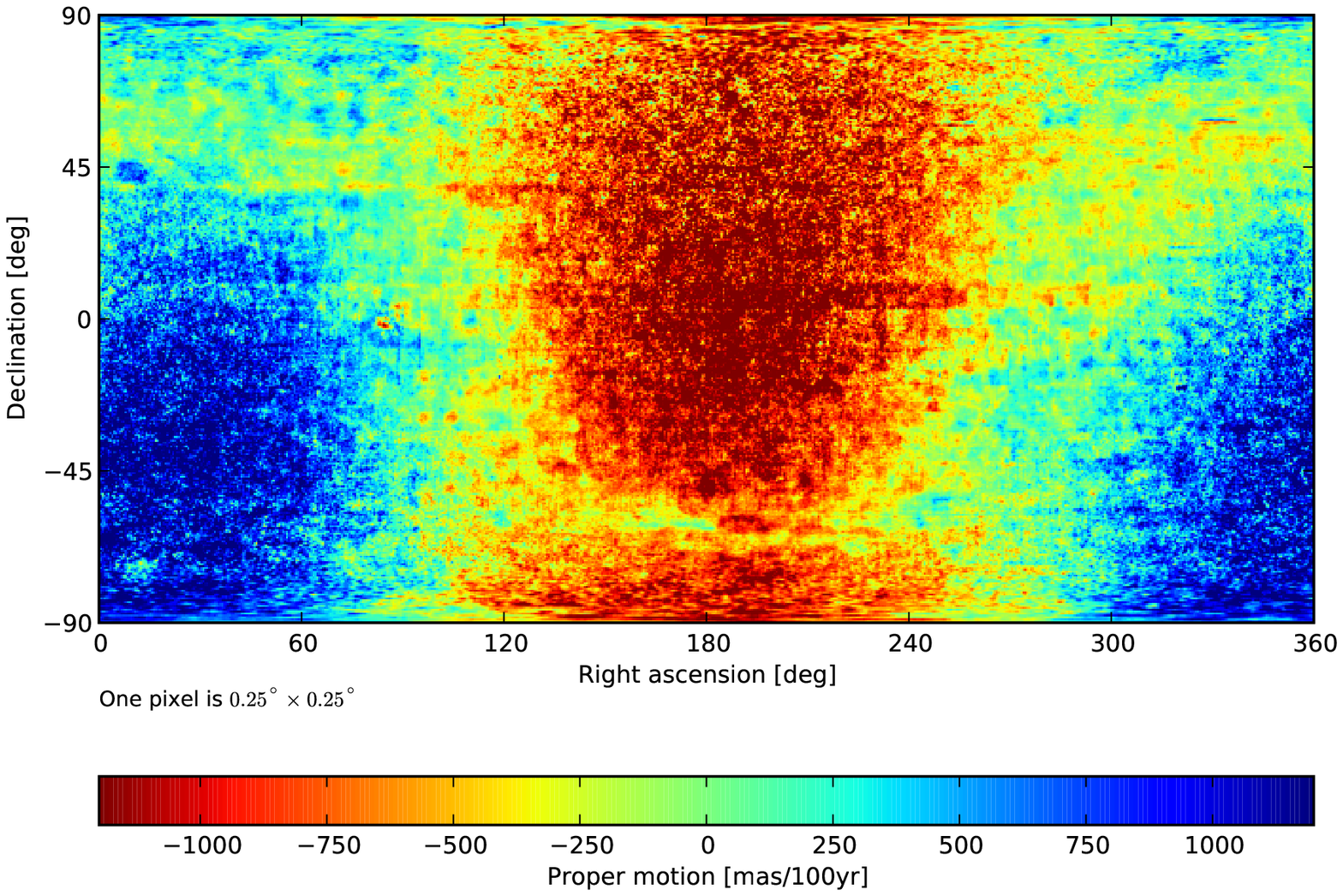}
\caption{Proper motions in right ascension $\mu_{\alpha}\cos\delta$
of the system of PPMXL represented by stars with
2MASS $K_s$-magnitudes between 12 and 13.
\label{muealfinal}}
\vspace*{0.1cm}

\figurehere
\includegraphics[bb= 36 230 550 580,angle=0,width=0.45\textwidth,clip]{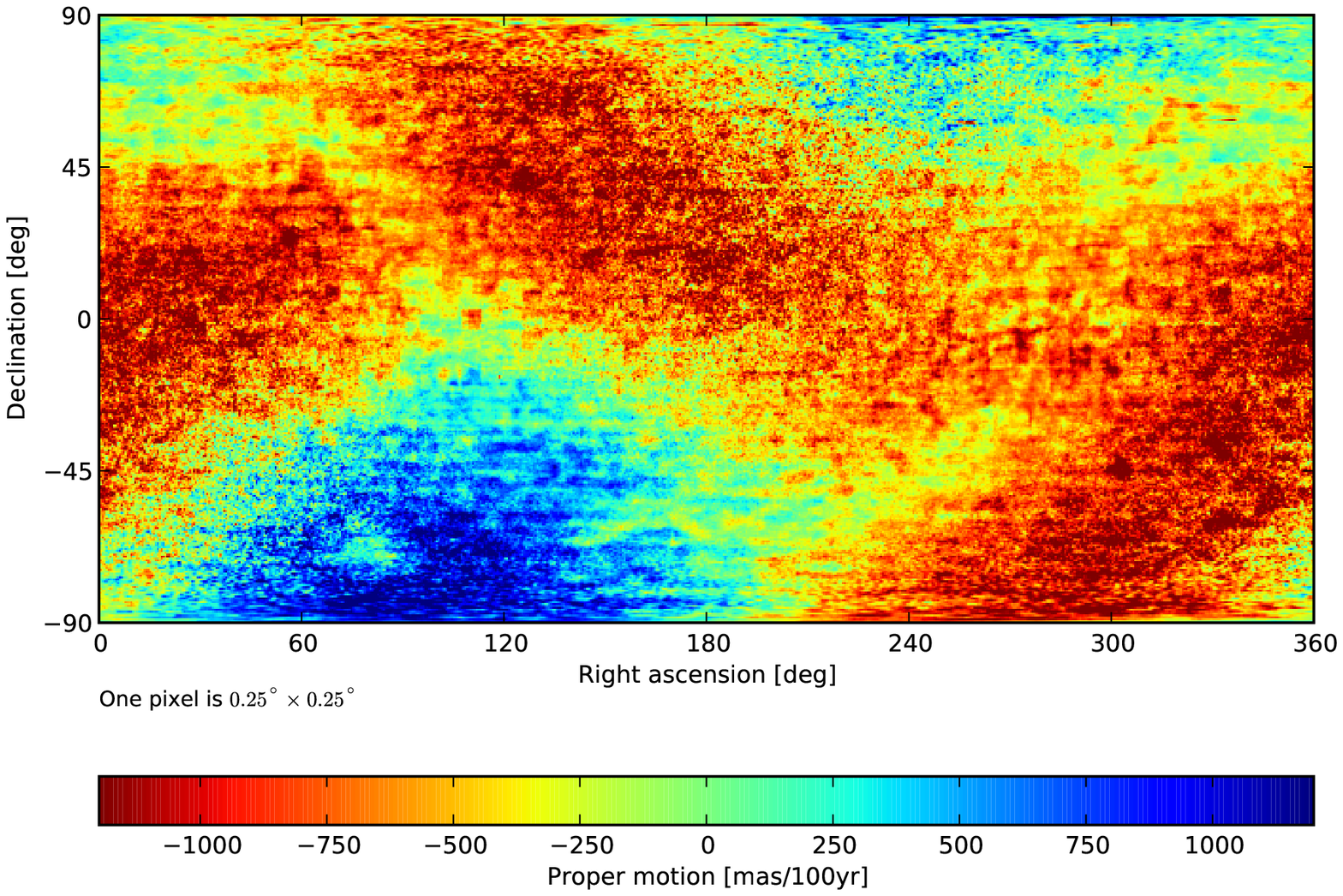}
\caption{Proper motions in declination $\mu_{\delta}$
of the system of PPMXL represented by stars with
2MASS $K_s$-magnitudes between 12 and 13.
\label{muedelfinal}}
\vspace*{0.1cm}

Qualitatively the faint stars show a similar pattern as
the bright ones (12 $< K_s <$ 13) in Fig. \ref{muedelfinal}. However, plate dependent
distortions still are not neglegible.
Empty areas display regions where the stars in PPMXL have no I-band observations in the 
range 18 $<$ I $<$ 19.
We attribute this to inhomogeneities in the USNO-B1.0 I-band
photometry,  which leads to an apparent underdensities or missing stars
compared to other fields.

In conclusion, the proper motions system at the faintest magnitudes is more uncertain than at bright ones.
A remedy could come from a sophisticated new reduction of all the survey plates where
care is taken to study and avoid all magnitude- and colour dependent effects. With Gaia
at the horizon, this effort probably is not warranted.   

\figurehere
\includegraphics[bb= 36 230 550 580,angle=0,width=0.45\textwidth,clip]{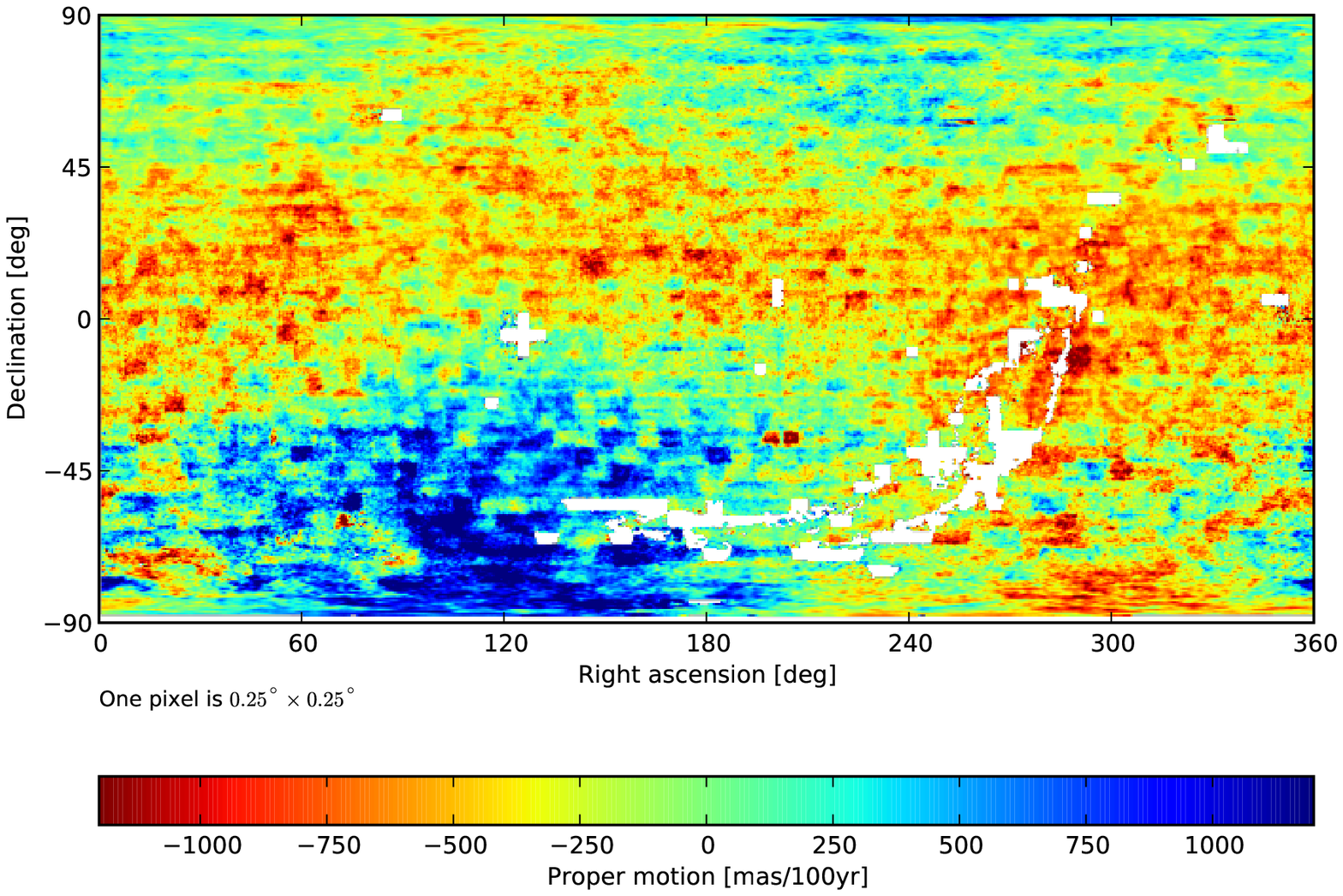}
\caption{Proper motions in declination $\mu_{\delta}$ of faint stars in PPMXL
(18 $<$ I $<$ 19). Qualitatively the faint stars show a similar pattern as
the bright ones (12 $< K_s  <$ 13 ) in Fig. \ref{muedelfinal}. However, plate dependent
distortions are not neglegible. Empty areas stand for regions where the stars in PPMXL
have no I-band observations in this
magnitude range.
\label{muedelfaint}}
\vspace*{0.1cm}

\subsection{The system of positions at 2000.0}~\label{possystem}

\begin{figure*}
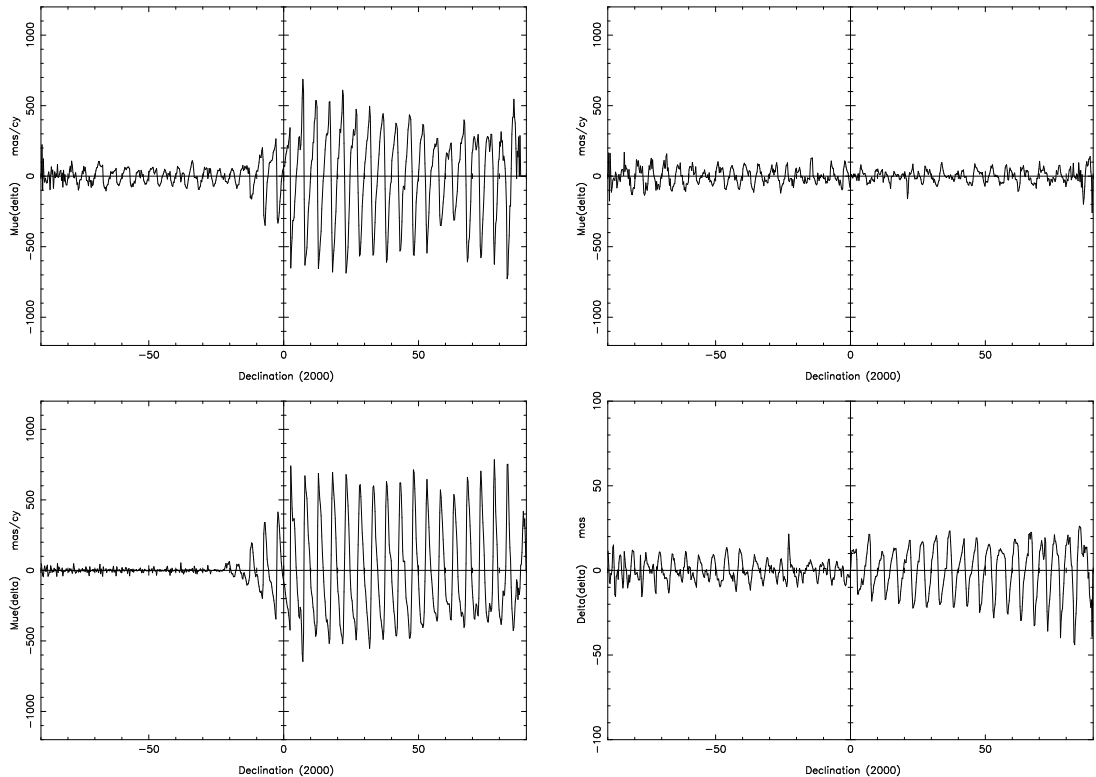

   \centering
    {\includegraphics[bb=60 10 580 750,angle=270,width=0.4\textwidth,scale=.99,clip]{muedelUdel.ps}}
    {\includegraphics[bb=60 10 580 750,angle=270,width=0.4\textwidth,scale=.99,clip]{muedelXLdel.ps}}\hfill
    {\includegraphics[bb=60 10 580 750,angle=270,width=0.4\textwidth,scale=.99,clip]{muedelXLmUdel.ps}}
    {\includegraphics[bb=60 10 580 750,angle=270,width=0.4\textwidth,scale=.99,clip]{ddelXLmUdel.ps}}\hfill
     \caption{The influence of the distorted UCAC3 proper motions
     in declination onto the
systematic accuracy of the compiled UCAC3 position at epoch 2000.0.
Large-scale ($ > 10^\circ$) variations have been filtered out by a high-pass (in frequency) filter.
Top left: the UCAC3 proper motions,
top right: the PPMXL proper motions,
bottom left: the difference in proper motions PPMXL-UCAC,
bottom right the difference in declination 2000.0 in the sense PPMXL-UCAC.
Units for proper motions are mas/cy, and
mas for the difference in declination. For explanation see text.  
     \label{pos2000} }
  \end{figure*}

Unlike the proper motions, the positions of PPMXL at J2000.0 can only be
referred to the ICRS by comparison with a catalog which is on ICRS.
The obvious choice would be UCAC3. However, UCAC3 does not publish the original observations
made in the years 1998 to 2004 with the CCD camera on the USNO astrograph
from CTIO and Flagstaff. In the published catalog the positions at 2000.0 are given,
which result from applying proper motions to the original positions. 
In the following we show that the positional system
of UCAC3 at 2000.0 is already corrupted by the systematically 
distorted proper motions even
over the few years between observation and 2000.0. Indeed, this can best be seen in Fig. \ref{pos2000}
for the declination system.
In section \ref{pmsystem} we showed the UCAC3 proper motions in declination
(see Fig. \ref{muedelU3S}).
If we average the proper motions over right ascension 
and let them pass through a high-pass filter (variations on scales larger than 10$^\circ$ are suppressed), 
we get on the northern sky 0$^\circ$ $ < \delta < $ 90$^\circ$ a wave-like behaviour with 
an amplitude of more than 5 mas/y,
a period of 5 degrees and phase 0 at the equator. A similar behaviour is
seen on the southern hemisphere but with a much smaller amplitude of only
1 mas/y. This is illustrated in the upper left panel of Fig. \ref{pos2000}. The upper right panel shows the same plot
(proper motions in declination)
for PPMXL.  On the northern hemisphere the difference is striking, whereas the coincidence
south of $\delta < -20^\circ $ is very good, as it should, because the system 
of PPMXL is strongly linked to UCAC3 in this area. This becomes much clearer in
the plot in the lower left, where we show the proper motion differences PPMXL - UCAC3 for 30 million stars
in common.

The lower right panel of Fig. \ref{pos2000} shows the differences in  positions at epoch 2000.0
between PPMXL and UCAC3.
Here we find again a 5$^\circ$ wave with an
amplitude increasing from the equator to the north pole (from 15 mas to 30 mas).
We note that the difference in proper motions 
(lower left panel) and the difference in declination (lower right panel) between PPMXL and UCAC3
are in anti-correlation; the explanation is given below.

At epoch 2000.0 the following equations hold (individually as well as averaged over right
ascension). Here, the subscript U stands for UCAC3, P for PPMXL.

\begin{equation}
 \delta_{U,2000} = \delta_{U,orig} + \mu_{\delta,U}\times(2000.0 - T_{U,orig})
\end{equation}
\begin{equation}
 \delta_{P,2000} = \delta_{P,orig} + \mu_{\delta,P}\times(2000.0 - T_{P,orig})
\end{equation}

Subtracting both equations
\begin{eqnarray}
\begin{tabular}{ l   }
$\delta_{P,2000} - \delta_{U,2000} = \delta_{P,orig}- \delta_{U,orig}+$\\
        
\hfill	     $    + \mu_{\delta,P}\times(2000.0 - T_{P,orig})+$ \\
	       
\hfill	     $   -  \mu_{\delta,U}\times(2000.0 - T_{U,orig})$\\ 
\end{tabular}	               
\end{eqnarray}

Note that in this magnitude range, $\delta_{P,orig}$ essentially coincides with the 2MASS declination,
because 2MASS has the highest weight and is almost at epoch 2000.0. 
The difference $\delta_{P,orig}- \delta_{U,orig} $ hence gives the difference between an original 2MASS
position and an original UCAC3 position where no proper motions (i.e. old epochs) are
involved. Both are reduced with Tycho-2, are almost coeval, and the
mean difference should be zero, at least not showing 5-degree waves.
Also, having filtered out the large scale effects, $\mu_{\delta,P}$ is
neglegibly small (smaller than 1 mas/y, Fig.\ref{pos2000}  upper right panel), so
we get:

\begin{equation}
\delta_{P,2000} - \delta_{U,2000} \approx -\mu_{\delta,U}\times (2000.0 - T_{U,orig}).
\end{equation}

For $\delta > 0^\circ$ we find that the position difference at 2000.0 is in phase with the 
UCAC3 proper motions as long
as $ (2000.0 - T_{U,orig})  <  0$, and in anti-phase otherwise. For the northern sky
$ (2000.0 - T_{U,orig})$ is negative because UCAC3 was observed from 2002 to 2004 from equator to pole,
exactly what we observe in Fig.\ref{pos2000} .
The increase in amplitude from equator to pole even reveals that the observations at the
pole came later than at the equator.
Although position differences of 15 to 30 mas are small, modern UCAC3
observations, accurate at epoch to 15 to 70 mas \citep{2004AJ....127.3043Z}
are considerably deteriorated systematically in only a few years.
In consequence, 
we could not use UCAC3 as reference catalog on ICRS for the position at 
epoch 2000.0.

The system PS2 of PPMXL consists of the proper motion system and the position system at J2000.0.
Like in the final step of the construction of the proper motion system (step 7), we chose PPMX
as representative of the positional system at epoch 2000.0,
so we added systematic differences PPMX-PS2 at epoch 2000.0 to PS2 on
a grid with 1$\times$1 deg$^2$ bins with a 3$\times$3 bin moving average filter
to achieve the final positional system of PPMXL at 2000.0.
In the case of stars having no 2MASS observations we proceeded analogously
to the proper motion system. This completes step 8.

\section{The final catalog}~\label{finalcat}

USNO-B1.0 contains more than a billion entries: stars and galaxies and 
a number of artefacts. \citet{2008AJ....135..414B} have detected
spurious entries in USNO-B1.0 that are caused by diffraction spikes and circular reflection
halos around bright stars in the original imaging data.
These defects, numbering some 24 million or 2.3 percent, were removed
using the data provided by \citet{2008AJ....135..414B}.

The final version of PPMXL contains some 900 million stars. We kept an entry from USNO-B1.0
whenever the maximum epoch difference between the observations 
was larger than 10 years. This somewhat arbitrary
choice was guided by the idea to formally derive proper motions even if a star
has only observations from 2MASS and the second epoch POSS, whereas no observations
from the first epoch POSS are available.
Because of this short epoch difference,
these stars have large mean errors of proper motions and they have to be used with care.

At its bright end, PPMXL is merged with PPMX according to the following scheme. 
The stars of PPMX were searched in PPMXL using a cone with 1.5 arcsec radius.
When no match was found, the resp. PPMX star was added to the catalog. This mainly
happened in the case of bright stars.
When a match has been found, the PPMX star is selected if the mean error of its
proper motion is smaller than that of the PPMXL star, and vice versa.
If a PPMX star is added to the catalog, all PPMXL matches within 1.5 arcsec are deleted.  

The photometric information from USNO-B1.0 is retained, as is the NIR photometry
from 2MASS if available. The data can be queried in
the VO and from http://vo.uni-hd.de/ppmxl, where a text dump is
available for download as well.  The catalog is also availabe at
CDS Strasbourg. 
 
\subsection{Properties}~\label{properties}

PPMXL contains 910,468,710 entries, including stars, galaxies and bogus entries. 412,410,368 of these are in 2MASS,
i.e. 2MASS is used to determine proper motions and the $J,H,K_s$ magnitudes are given
in the catalog. 6,268,118 stars are taken from PPMX, so PPMXL aims to be complete from
the brightest stars down to about 20th magnitude in V.

The covariance matrix obtained in the least-squares adjustment in section \ref{lsqadj} gives
(per coordinate and per star) the mean epoch, the mean error of position at mean epoch and
the mean error of proper motions. All these quantities are published in the catalog.

Mean errors of the positions at the reference epoch 2000.0 can be computed star by star.
On average, the mean errors of position 2000.0 are between 80 and 120 mas if 2MASS astrometry is available,
and range from 150 mas to 300 mas else.

\begin{figure*}
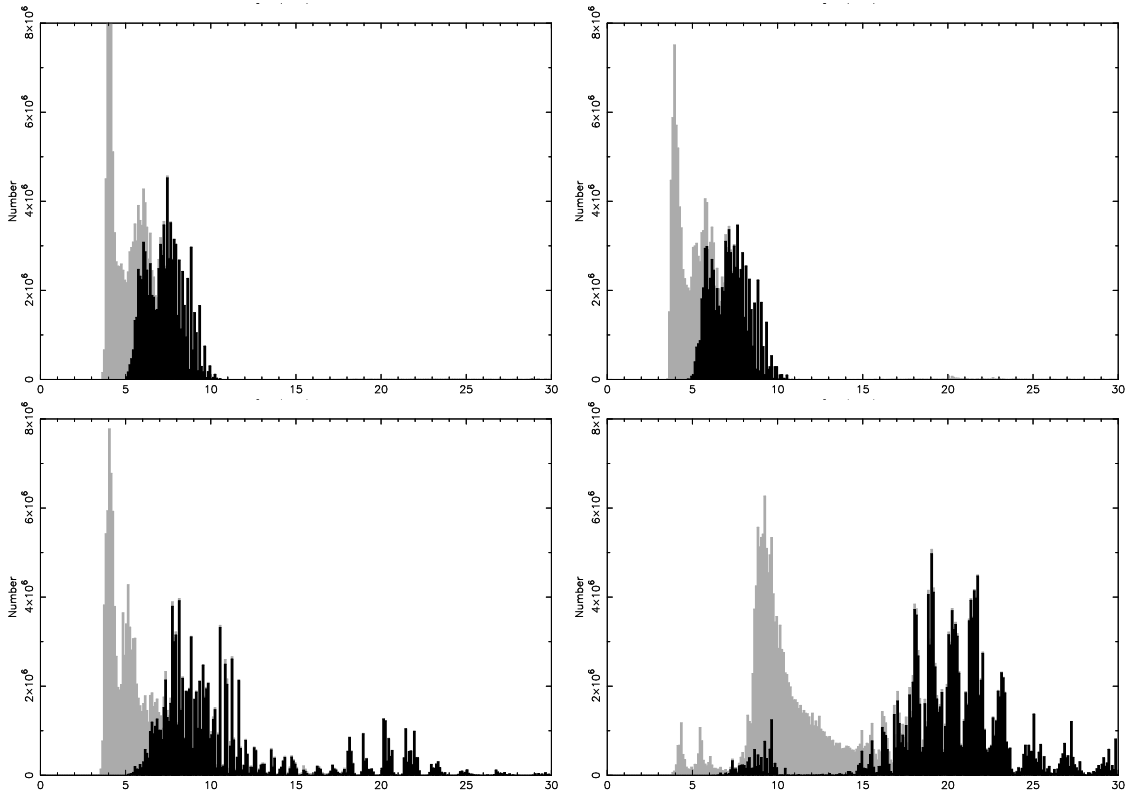

   \centering
    {\includegraphics[bb=57 34 553 737,angle=270,width=0.4\textwidth,scale=.99,clip]{smueap30p90.ps}}
    {\includegraphics[bb=57 34 553 737,angle=270,width=0.4\textwidth,scale=.99,clip]{smueap00p30.ps}}\hfill
    {\includegraphics[bb=57 34 553 737,angle=270,width=0.4\textwidth,scale=.99,clip]{smueam30m00.ps}}
    {\includegraphics[bb=57 34 553 737,angle=270,width=0.4\textwidth,scale=.99,clip]{smueam90m30.ps}}\hfill
     \caption{
Distribution of the formal mean error of a proper motion component (here $\mu_{\alpha} cos\delta$)
in units of $mas/y$. Shown are the 
four quarters of the sky: top left +90$^\circ$ $ > \delta \geq $ +30$^\circ$,
top right +30$^\circ$ $ > \delta \geq $ 0$^\circ$,
bottom left 0$^\circ$ $ > \delta \geq $ -30$^\circ$,
bottom right -30$^\circ$ $ > \delta > $ -90$^\circ$. The envelope is the distribution of all stars,
stars shaded in grey are those having 2MASS observations, stars in black do not have 2MASS observations.   
     \label{sigmamue} }
  \end{figure*}

The statistics of the mean errors of the proper motions resembles the
observational history rather than the poorer signal-to-noise at fainter magnitudes. In Fig.\ref{sigmamue}
we present the distributions of the mean errors of the proper motions in 4
declination zones. The following can be drawn from this figure.
Including the measurements from 
2MASS yields a considerable improvement, both because of its good accuracy and its
recent epoch. The latter effect is most pronounced on the southern sky. 
At $ \delta < -30^\circ$, the proper motions without 2MASS are of poor quality,
because the first epoch is much later than in the other zones,
and hence the epoch difference is rather small. 2MASS improves the situation, but a contemporary
(2010) survey such as the Sky Mapper Southern Sky survey
\citep{2007PASA...24....1K} will give a considerable improvement.

PPMXL is a catalog which is nominally on the ICRS system. It is linked to
the Hipparcos catalog, the optical representation of the ICRS, via Tycho-2
and PPMX. A word about the inertiality of PPMXL is therefore appropriate.
The uncertainty of a residual rotation of Hipparcos itself is 0.25 mas/y
\citep{1997A&A...323..620K}. This is a global quantity, on smaller
scales the uncertainty is larger.
On a typical field of the sky of one square degree we
find about 3 faint Hipparcos stars with an rms error of
the proper motion of, say, 1.7 mas/y each.  Therefore, their ``average 
motion'' has a mean error of roughly 1 mas/y, a figure
representative of the uncertainty of the deviation of Hipparcos from
a truly inertial system on a one degree scale.
The actual value of such a deviation can only be determined with the results from Gaia
or other space missions with limiting magnitude deeper than Hipparcos
such as JMAPS and, perhaps, nano-JASMINE.
Also, full-sky block-adjustment procedures applied to old and new surveys can help. 
The two intermediate steps from Hipparcos to PPMXL introduce additional systematic errors
which cannot be estimated rigorously. It is therefore not unreasonable to state that
the absolute proper motions given in PPMXL have an underlying systematic uncertainty
of at least 1 to 2 mas/y, which is small compared to the random error for the vast
majority of PPMXL stars. The mean motion of an ensemble of stars, however,
cannot be determined better than the 1 to 2 mas/y mentioned above. These arguments, of course,
hold similarly to Tycho-2, PPMX and the UCAC series of catalogs. 




\subsection{Caveats}~\label{Caveats}
{\em Stars with $n_{obs} = 2$ or with {\rm 'P'} flags}:
About 146 million stars (or 16\%) have proper motions based on 2 observations only;
and some 77 million (or 8.4\%) carry the P flag defined in section \ref{lsqadj} denoting
bad $\chi^2$ sums of the residuals after the lsq adjustment. Both cases concentrate
towards the edges of plates and in dense regions of the southern sky.

{\em Magnitude system}:
We made no attempt to recalibrate the USNO B1.0 magnitude system.
There are discrepancies in the magnitude system from field
to field and from early to late epoch. In principle, the magnitudes can be calibrated 
using the Guide Star Photometric Catalog
\citep{2001A&A...368..335B}.

{\em Double entries}: Cross-matching with 2MASS has been performed using a cone of 3 arcsec
radius. Given the spatial resolution on the Schmidt-plates underlying USNO-B1.0 only a single match between a 2MASS
and a USNO-B1.0 entry should occur. However, on the northern hemisphere a single match
was found in 93.6$\%$ of all cases, on the southern hemisphere in only 88.7$\%$.
There are also triple and multiple matches, but their number is smaller than 0.03$\%$ north,
0.16$\%$ south.
The number of doubles and multiples strongly increases at plate boundaries and in
dense regions on the southern sky (LMC, inner Galactic plane).
No complete auto-crossmatch has been
made with PPMXL, but extrapolating the matches with 2MASS to the full 900 million objects,
we estimate that about 90 million (10$\%$) are false doubles or multiples from USNO-B1.0.
Turning the cross-match around we also found double matches of 2MASS stars with an USNO-B1.0 star.
This amounts to 1.5$\%$ of all cases. So, 2MASS has between 6 and 7 million doubles.

{\em ''High proper motion'' stars}:
There is a huge number of stars with high proper motions, e.g. on the northern
hemisphere about 24.5 million objects have proper motions
larger than 150 mas/y. The vast majority of them must be fakes; 
a practically flat proper motion distribution function between 130 and 430 mas/y
is a hint to this.
Also, the LSPM-NORTH Catalog \citep{2005AJ....129.1483L} lists only some 61000 stars
on the northern hemisphere
with proper motions larger than 150 mas/y and claims to be complete
to V = 19. An attempt to solve the problem
with these large proper motions (already inherent in USNO-B1.0) is far beyond
this paper; we only note in passing that many cases
occur among the above-mentioned doubles or multiples.

\section{Correction tables for observations based upon USNO-B1.0}

Observers using USNO-B1.0 for the reduction of their CCD frames get
positions of their targets which are not on the ICRS. Such positions
can neither be used to derive inertial stellar proper motions, nor
should they be used for in orbit determinations
in the case of solar system bodies.
According to \citet{2009DDA....40.1704C}, there are millions of minor planet positions
based on USNO-B1.0 in recent years.

To aid in reducing these observations to ICRS we present systematic correction tables
from USNO-B1.0 to PPMXL for positions at epoch 2000.0 and for proper motions.
The tables give the means of the four quantities in circles of radius
$\sqrt{2}/2$ degrees around the centers of 360 by 180 spherical squares covering
the sky.
The application of these tables is straighforward. Suppose you have an observation
($\alpha,\delta$) based on USNO-B1.0 at epoch T (in years). For this $\alpha,\delta$
the tables give four quantities $\Delta\alpha, \Delta\delta, 
\Delta\mu_{\alpha}\cos\delta, \Delta\mu_{\delta}$ in the sense PPMXL - USNO-B1.0. Then the conversion to ICRS
is given by

\begin{eqnarray}
\begin{tabular}{ l   }
 $ \alpha_{{\rm ICRS}} = \alpha +\Delta\alpha +(\Delta\mu_{\alpha}\cos\delta)/\cos\delta\times(T - 2000.0)$ 
\end{tabular}
\end{eqnarray}
\begin{equation}
\delta_{{\rm ICRS}} = \delta + \Delta\delta+ \Delta\mu_{\delta}\times(T - 2000.0) 
\end{equation}

The correction tables can be used or downloaded from http://vo.uni-hd.de/ppmxl.
In applying these formulae note that they are not rigorous near the poles, and also
that a star can cross the $\alpha = 0$ border when one applies the corrections in right ascension.
We will also
deliver correction tables for USNO-A2.0, UCAC2 and UCAC3 and 2MASS
on the same server. Note that 2MASS is an observational catalog, literally
speaking with proper motions $\equiv 0$, observed in the years 1997 and 2001.
So, in 2010 {\em systematic} offsets may already reach 150 mas (15 mas/y$\times$10 years).
All the correction tables are created under the implicit assumption
that PPMXL has no magnitude dependent systematics.  







\acknowledgments

This research made use of the facilities of the German Astrophysical
Virtual Observatory (GAVO).
This publication refers to
observations from the ESA Hipparcos satellite.
It makes use of data products from the Two Micron All Sky Survey,
which is a joint project of the University of Massachusetts and the
Infrared Processing and Analysis Center/California Institute of Technology,
funded by the National Aeronautics and Space Administration and the National Science Foundation.
This research has made use of the SIMBAD database,
operated at CDS, Strasbourg, France.
We also acknowledge the
institutions that funded the sky surveys on which USNO-B1.0 is based upon.

We are grateful to Nina Kharchenko for quality checks in regions of a few
open clusters.  We thank Dave Monet for supplying a complete list of
field epochs of USNO-B1.0

\clearpage

\end{document}